\documentclass{article}

\usepackage{PRIMEarxiv}
\usepackage{amsmath} 
\usepackage[utf8]{inputenc} 
\usepackage[T1]{fontenc}    
\usepackage{hyperref}       
\usepackage{url}            
\usepackage{booktabs}       
\usepackage{amsfonts}       
\usepackage{nicefrac}       
\usepackage{microtype}      
\usepackage{lipsum}
\usepackage{fancyhdr}       
\usepackage{graphicx}       
\graphicspath{{media/}}     

\title{Modeling of Asymptotically Periodic Outbreaks: \\
a long-term SIRW2 description of Covid-19? }

\author{
 Alex Viguerie \\
  Division of Mathematics\\
  Gran Sasso Science Institute\\
  Viale Francesco Crispi 7 \\
  L'Aquila, AQ 67100, Italy \\
  \texttt{alexander.viguerie@gssi.it} \\
  \And
 Margherita Carletti \\
  Department of Pure and Applied Sciences\\
  University of Urbino C. Bo\\
  Piazza della Repubblica 13\\
  Urbino (Pu) 61029, Italy\\
  \texttt{margherita.carletti@uniurb.it} \\
  \And
  Alessandro Veneziani \\
  Department of Mathematics\\
  Department of Computer Science\\
  Emory University\\
  400 Dowman Drive \\
  Atlanta, GA 30322, USA \\
  \texttt{avenez2@emory.edu} \\
  \And
 Guido Silvestri \\
Yerkes Division of Microbiology and Immunology\\ Yerkes National Primate Research Center\\
Yerkes Nonhuman Primate Genomics Core\\
Emory University \\
Atlanta, GA 30329, USA \\
  \texttt{gsilves@emory.edu} \\
}

\begin{document}
\maketitle

\begin{abstract}
As the outbreak of COVID-19 enters its third year, we have now enough data to analyse the behavior of the pandemic with mathematical models
over a long period of time. The pandemic alternates periods of high and low infections, in a way that sheds a light on the
nature of mathematical model that can be used for reliable predictions. The main hypothesis of the model presented here
is that the oscillatory behavior is a structural feature of the outbreak, even without 
postulating a time-dependence of the coefficients. As such, it should be reflected by the presence of 
limit cycles as asymptotic solutions. This stems from the introduction of (i) a non-linear waning immunity
based on the concept of immunity booster (already used for other pathologies); (ii) a fine description of the compartments
with a discrimination between individuals infected/vaccinated for the first time, and individuals already infected/vaccinated, undergoing to 
new infections/doses. 
We provide a proof-of-concept that our novel model is capable of reproducing 
long-term oscillatory behavior of many infectious diseases, and, in particular, 
the periodic nature of the waves of infection. 
Periodic solutions are inherent to the model, and achieved without changing parameter values in time. 
This may represent an important step in the long-term modeling of COVID-19 and similar diseases, as the natural, unforced behavior of the solution shows the qualitative characteristics observed during the COVID-19 pandemic.
\end{abstract}

\section{Introduction}

\begin{figure}[h]\centering
  \includegraphics[width=\textwidth]{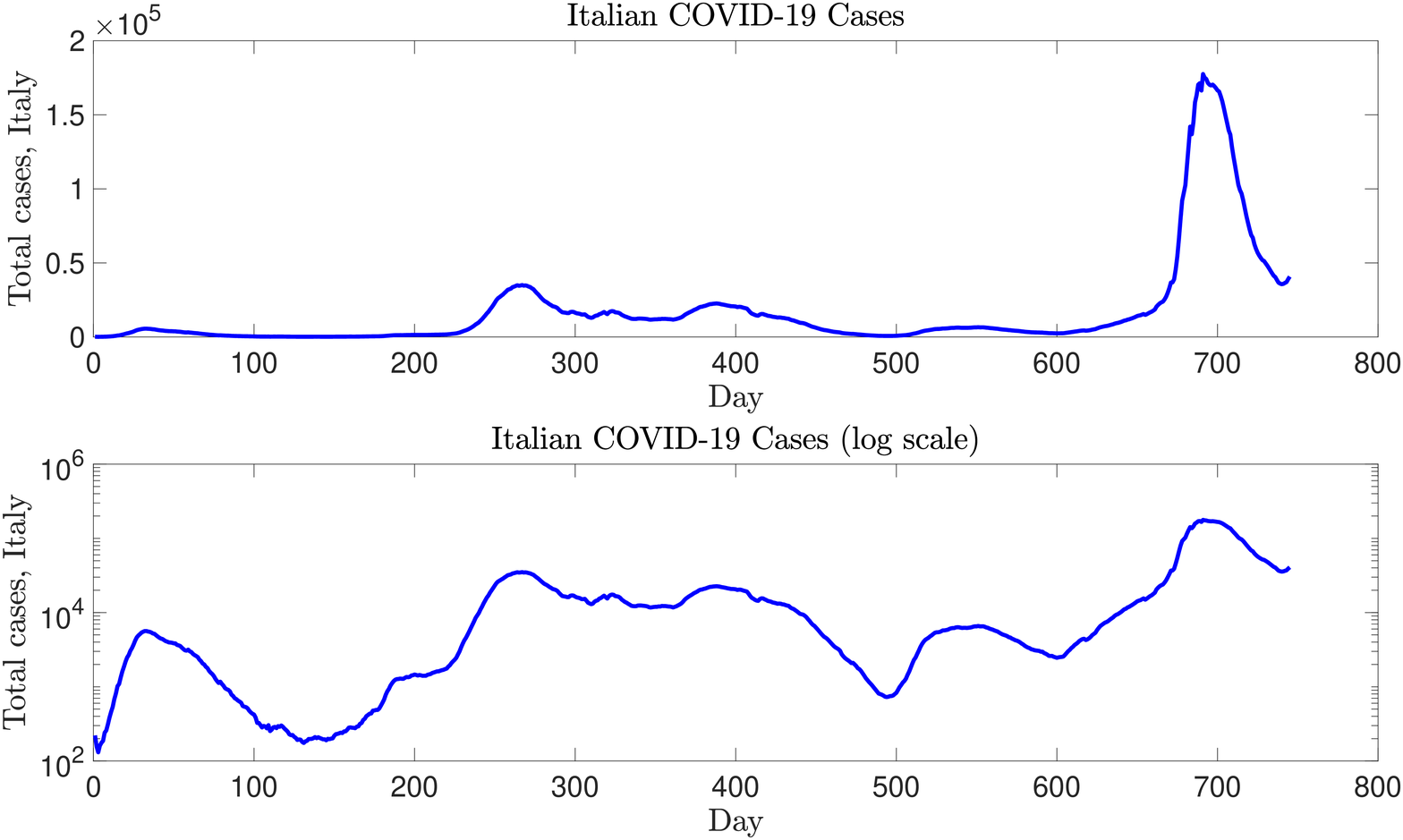}
 \noindent \includegraphics[width=\textwidth]{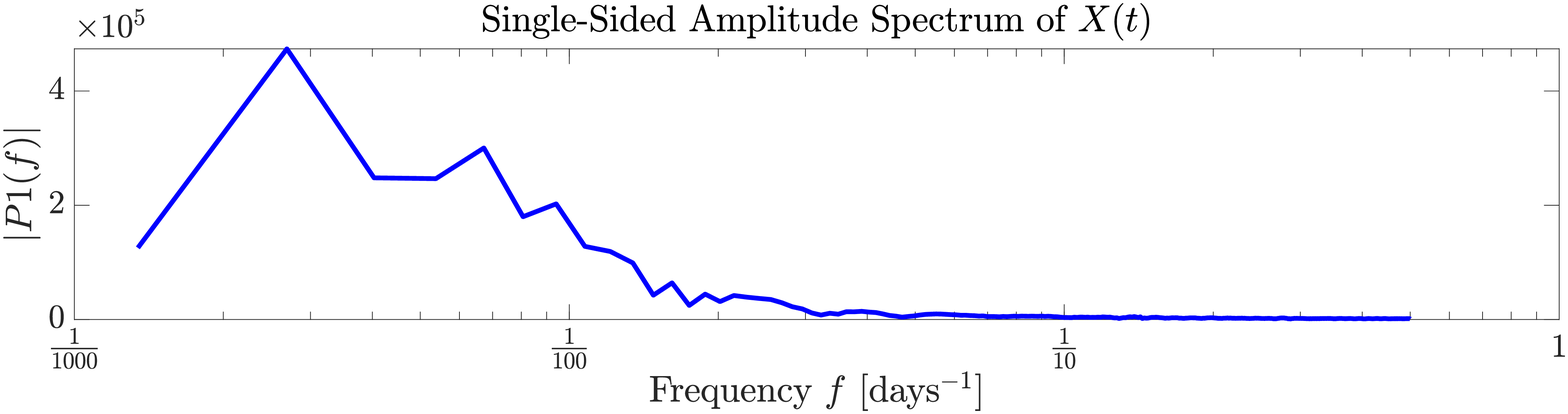}
 \caption{Infected population in Italy over the time range Feb 24 2020- Mar 9 2022. Top: linear scale, center: log-scale. We observe a clear oscillatory, periodic dynamics.
 Bottom: Fourer spectral analysis showing the dominant frequencies (1/370 days and
 1/148 days).}\label{fig:infItaly}
\end{figure}

It has been two years now that the world experienced the COVID-19 pandemic. We have enough data to identify some patterns in the evolution of the outbreak over time.
In particular, in many situations it is evident that the outbreak features an oscillatory behavior, showing an alternating behavior of high and low infected populations. For example, in Fig. \ref{fig:infItaly}, we plot the infected curve in Italy (source: \cite{ItalianPC}). 
We also plot a Fourier spectral analysis of the data, showing that the dominant frequencies are 
around 1 year$^{-1}$ and three months$^{-1}$.
Whether this is the effect
of a seasonal infection-rate, government measures, or a structural nature of the pandemic is currently an open question.

The calibration of mathematical models with a predictive purpose can be widely educated
by the answer to this question. If we assume that the periodicity is a structural feature of the pandemic, a mathematical model should reflect this in the nature of its asymptotic solutions.
On the contrary, if the periodicity is the result of seasonality, or government responses to the pandemic (such as lockdowns and masking mandates), the model should incorporate this effect in its parameter calibration. 

In the majority of work modeling COVID-19, this latter assumption is employed. Such works typically adopt classical compartmental models like SIR (Susceptible-Infected-Recovered, see \cite{kermack1927contribution, breda2012formulation} for detailed discussion), and its derivations. These models may lead to excellent short-term predictions; however, as they do not have periodic solutions as asymptotic time limit, they will ultimately fail to yield a periodic solution
when using time-independent parameters - they do not have periodic asymptotic solutions \cite{hethcote2000mathematics}. In order to recover the observed periodic behavior, then, such models must employ time-dependent parameter fitting, generally attributing the periodicity to underlying changes in population behavior, seasonality, government measures, and other factors \cite{viguerie2021simulating, grave2021assessing, zhang2021integrated, parolini2021suihter, bhouri2021covid}. In contrast, in order to recover a truly periodic solution without regular re-parameterization, the presence of limit-cycles 
becomes a mandatory aspect of long-term modeling. 

It is worth noting that the two different possible answers (nature of the problem vs. time-dependence of the parameters) are not mutually exclusive.
Looking at the spectral analysis of the Italian data,
we argue that it may be a combination of different effects
on long and short time scales.


In this paper, we focus on a model where the wavy pattern is primarily  
intrinsic to the infection \textit{qualitatively}, acknowledging that behavioral, seasonal, and other external aspects
(e.g., higher infectivity in Winter, lower in Summer, different infectivity of different variants, lockdowns)
impact the pattern \textit{quantitatively}.
 
Our starting point is the concept of \textit{waning immunity} advocated for other pathologies \cite{jardon2021geometric}.
Waning Immunity has been already considered for COVID-19, see, e.g., \cite{anggriani2022mathematical,crellen2021dynamics,altmann2021waning,dolgin2021covid} 
with different mathematical descriptions.
The basic idea is that the immunity induced by recovery or vaccination declines over time.
This was largely demonstrated by the cases of reinfection/infection-after-vaccination
documented for the COVID-19 outbreak. However, the reduction of the immunity does not follow a simple pattern, because there is an \textit{immunity booster effect}:
an individual with a declining immunity exposed to an infected person may get a beneficial effect, reinforcing (or boosting) her/his immune system.
This leads to a model sometimes called SIRW, where the compartment W (Waning) added to the classical SIR denotes the ``weakly immune'' individuals that 
may go back to the Susceptible compartment or get an immunity booster by contact with infected or vaccination.
It has been demonstrated by a geometric approach \cite{jardon2021geometric} that this model does have limit cycles for particular combinations of the parameters.
One of the important aspects of the parameter calibration is the ``multiscale-in-time'' nature of the model, where the natural growth of the population
(according to a Malthusian model) is, in fact, a slower process compared to the infection/vaccination rate.

In this brief note, we propose a new extension of this model. We further stratify all four subpopulations of the SIRW into ``never previously exposed'' and ``previously exposed" subpopulations. The rationale is that, while a vaccinated or previously infected individual may lose their immunity from infection over time, some baseline protection is nonetheless maintained; the illness is less likely to be severe and is more likely to pass quickly. A similar concept in the case of pertussis (with no immunity booster) was advocated in \cite{van2000waning}. 
As the model is formally a duplication of the four SIRW compartments (plus the Deceased), we call this model SIRW2 (SIRWsquare).

The main purpose of this communication is to provide a proof-of-concept that the SIRW2 model can actually describe the wavy behavior  even in the absence of 
a time-dependent calibration of the parameters. The presented work presents preliminary results
that suggest that this model can provide a reliable long-term description of the COVID-19 outbreak.

The document is organized as follows. We first introduce the waning immunity 
and explain the SIRW model (studied in \cite{jardon2021geometric}), in order to describe the immune-boosting process and provide the motivation behind the SIRW2 model. We then introduce the full SIRW2 model, providing the model equations, a description of the dynamics they are intended to model, and a complete list of the relevant parameters and their purpose. We then numerically demonstrate with constant  parameters  that this model may in principle reproduce wave-like dynamics. 
We conclude with a series of perspectives on future steps in terms of both theoretical analysis and
validation.
We aim at testing the hypothesis that the periodic dynamics of COVID-19 pandemic is intrinsic to the nature of the infection, so to be properly described (and quantitatively predicted) by the SIRW2 model.


\section{The Mathematical Model}

\subsection{The SIRW model}
The SIRW model shown in \cite{jardon2021geometric} considers the \textit{susceptible} $s$, \textit{infected} $i$, \textit{recovered} $r$, and \textit{waning} $w$ compartments and reads as follows:

\begin{align}\label{eqn:sirw_s}
d_t s &= -\beta s i + \kappa w, \\ \label{eqn:sirw_i}
d_t i &= \beta s i - \phi i, \\ \label{eqn:sirw_r}
d_t r &= \phi i - \xi r + \chi i w, \\ \label{eqn:sirw_w}
d_t w &= \xi r - \chi i w - \kappa w.
\end{align}
 A full list of the variables and parameters in \eqref{eqn:sirw_s}-\eqref{eqn:sirw_w} is provided in Table \ref{tab:SIRW_Parameter_Table}. 
\par The model \eqref{eqn:sirw_s}-\eqref{eqn:sirw_w} describes the passage of individuals throughout the four compartments. A susceptible individual ($s$), upon contact with an infected individual $i$, may become infected (the rate of which is controlled by the contact rate $\beta$ and proportional to the product $si$), moving into the compartment $i$. Infected individuals recover with a rate $\phi$, moving into the $r$ compartment, where they acquire immunity from infection. At a rate of $\xi$, recovered individuals begin to lose their immunity, entering the waning compartment $w$.
\par While the immunity is waning, two things can happen:
\begin{enumerate}
    \item[(i)] The waning individual can come into contact with an infected individual again, with contact rate $\chi$, boosting their immunity, returning them to the $r$ compartment, or;
    \item[(ii)] The waning individual, with no additional contact with the virus, loses their immunity and returns to the $s$ compartment at rate $\kappa$.
\end{enumerate}
In \cite{jardon2021geometric}, the model \eqref{eqn:sirw_s}-\eqref{eqn:sirw_w} was shown to admit periodic-orbit solutions in time, suggesting a correspondence between the modeled immune-boosting process and the possible oscillatory dynamics. Thus, we believe this model, and the underlying immune-boosting mechanism, may provide a key towards a more realistic model of COVID-19, and the immune process generally.

\begin{table}
\begin{center}
\begin{tabular}{ |c|c|c| } 
\hline
Parameter/variable &  Name  &  Units   \\
\hline\hline

$s$ & Susceptible individuals & Persons \\ \hline
$i$ & Infected individuals  & Persons \\ \hline
$r$ & Recovered individuals  & Persons \\ \hline
$w$ & Waning-immunity individuals & Persons \\ \hline
$m$ & Deceased individuals & Persons \\ \hline
$\beta$ & Contact rate between $s$ and $i$ & Persons$^{-1} \cdot $ Days$^{-1}$  \\ \hline
$\phi$ & Recovery rate for & Days$^{-1}$  \\ \hline
$\xi$ & Loss-of-immunity initiation rate & Days$^{-1}$  \\ \hline
$\chi$ & Immunity boosting-contact rate & Persons$^{-1} \cdot $ Days$^{-1}$   \\ \hline
$\kappa$ & Loss-of-immunity rate & Days$^{-1}$  \\ \hline
\end{tabular}
\caption{Relevant variables, parameters, parameter names, and units for the systems \eqref{eqn:sirw_s}-\eqref{eqn:sirw_w}.}
  \label{tab:SIRW_Parameter_Table} 
\end{center}
\end{table}

\subsection{The SIRW2 model}

While the more detailed immunological process described in model \eqref{eqn:sirw_s}-\eqref{eqn:sirw_w} provides a sensible starting point for the modeling of COVID-19,
preliminary numerical simulations show that the oscillatory behavior is monotonically damped,
a circumstance that is not corresponded by data.
In fact, the model \eqref{eqn:sirw_s}-\eqref{eqn:sirw_w} does not consider vaccinations and does not consider mortality from the disease. The addition of several new terms for vaccination and mortality, as well as an additional compartment for deceased is sufficient for their inclusion.

The more important modification necessary, however, is that the loss of immunity is not
a complete and total elimination of all immunity.
In reality, individuals maintain some baseline of immunological protection, 
as explained above, in the form of, for instance, Memory B- cells \cite{laidlaw2022germinal}. These protections both protect against severe disease, and lead to a faster recovery process upon reinfection, when compared to infection in an individual with no prior immunological protection. 
\par To this end, we introduce an extension of \eqref{eqn:sirw_s}-\eqref{eqn:sirw_w} incorporating these crucial effects: vaccination, disease mortality, and distinction between populations with and without prior immunological exposure. To model vaccination, we add a vaccination term, which can both grant and boost immunity (i.e. enable the movement $s \to r$ and $w \to r$). For the addition of mortality, we add another compartment, denoted $m$ (for `mortality'), to track deceased individuals. Finally, we stratify the populations based on whether or not they have prior immunological protection, either from vaccination or prior infection.
We also incorporate a Malthus demographic independent of the disease into the model.
The extended model, which we denote by SIRW2 (SIRW-squared) model, reads:

\begin{align}\label{eqn:s_u}
d_t s_u &= -\beta_{u}^u s_u i_u - \beta_{u}^e s_u i_e - \alpha s_u + \psi n- \rho s_u, \\ \label{eqn:i_u}
d_t i_u &= \beta_{u}^u s_u i_u + \beta_{u}^e s_u i_e - \phi_u i_u - \gamma_u i_u - \rho i_u, \\ \label{eqn:r_u}
d_t r_u &= \phi_u i_u + \alpha s_u - \xi_u r_u + \chi_u^e w_u i_e + \chi_u^u w_u i_ u + \alpha w_u - \rho r_u, \\ \label{eqn:w_u}
d_t w_u &= \xi_u r_u -  \chi_u^e w_u i_e - \chi_u^u w_u i_ u - \kappa_u w_u - \alpha w_u - \rho w_u, \\ \label{eqn:s_e}
d_t s_e &= -\beta_{e}^u s_e i_u - \beta_{e}^e s_e i_e + \kappa_u w_u + \kappa_e w_e - \alpha s_e -  \rho s_e, \\  \label{eqn:i_e}
d_t i_e &= \beta_{e}^u s_e i_u + \beta_{e}^e s_e i_e  - \phi_e i_e - \gamma_e i_e - \rho i_e, \\ \label{eqn:r_e}
d_t r_e &= \phi_e i_e - \xi_e r_e + \chi_e^e w_e i_e + \chi_e^u w_e i_ u + \alpha w_e - \rho r_e,  \\ \label{eqn:w_e}
d_t w_e &= \xi_e r_e - \chi_e^e w_e i_e - \chi_e^u w_e i_ u - \kappa_e w_e - \alpha w_e - \rho w_e, \\ \label{eqn:m}
d_t m &= \gamma_u i_u + \gamma_e i_e.
\end{align}




Here, $n=  s_u+i_u+r_u+w_u + s_e+i_e+r_e+w_e $, is the size of the living population. The subscript $u$ indicates `unexposed', meaning this class of individuals has not been previously infected with, or vaccinated against, the disease. The subscript $e$ indicates `previously exposed', meaning that this class of individuals has been infected previously, or vaccinated against, the disease. 
\par Verbally, the model \eqref{eqn:s_u}-\eqref{eqn:m} describes the population of persons who have never been infected with the disease $s_u$, who may become infected via contact with either a first-time infected ($i_u$) or reinfected ($i_e$) person. After moving to the first-infection compartment $i_u$, a person may recover (with rate $\phi_u$) or die (rate $\gamma_u$). Assuming recovery, the movement is into the compartment $r_u$, indicating recovery with immunity. Alternatively, an individual may be vaccinated (with rate $\alpha$) and move directly from $s_u$ to $r_u$ without any previous infection. 
\par Once in the the recovered state $r_u$, an individual begins to lose immunity, and, at a rate of $\xi_u$, the individual moves to the waning-state $w_u$. In the waning-state, a few things can happen:
\begin{enumerate}
    \item[(i)] An individual can come into contact with an infected individual (either $i_u$ or $i_e$, with contact rate $\chi_u,\,\chi_e$ respectively, boosting the immunity and returning to the compartment $r_u$;
    \item[(ii)] An individual can receive a vaccine booster (rate $\alpha$), also boosting their immunity and returning to $r_u$;
    \item[(iii)] In the absence of additional contact with the virus, in the form of exposure to an infected individual or a vaccine booster, the immunity can be lost with a rate $\kappa_u$. 
\end{enumerate}
\par In this last case, the switch is to the compartment $s_e$. From there, the model progression works similarly to in the first-infection (sub-scripted with $u$) case, in the movement between the $i_e$, $r_e$, and $w_e$ states. However, the previously-exposed infection progression differs from the first-time infection progression in that the relevant parameters are different. For most diseases we postulate that:
\begin{enumerate}
    \item[(iv)] The mortality rate is substantially lower: $\gamma_e$ << $\gamma_u$;
    \item[(v)] The recovery rate is substantially higher: $\phi_e$ >> $\phi_u$.
    \end{enumerate}
This is in line with observation and general consensus. Our model evidence also suggests that, perhaps:
    \begin{enumerate}
\item[a)] Immunity is lost more quickly after reinfection: $\xi_e$ > $\xi_u$; 
\item[b)] however, it is also boosted more easily among those with prior immunity: $\chi_e^e > \chi_u^e$, $\chi_e^u > \chi_u^u$.
\end{enumerate}

A full list and short description of the different variables, parameters, and their units is given in Table \eqref{tab:1DParametertTable}.

\begin{table}
\begin{center}
\begin{tabular}{ |c|c|c| } 
\hline
Parameter/variable &  Name  &  Units   \\
\hline\hline

$s_{u,e}$ & Susceptible individuals (no prior immunity, prior immunity) & Persons \\ \hline
$i_{u,e}$ & Infected individuals (no prior immunity, prior immunity) & Persons \\ \hline
$r_{u,e}$ & Recovered individuals (no prior immunity, prior immunity) & Persons \\ \hline
$w_{u,e}$ & Waning-immunity individuals (no prior immunity, prior immunity) & Persons \\ \hline
$m$ & Deceased individuals & Persons \\ \hline
$\beta_u^{u,e}$, $\beta_e^{u,e}$ & Contact rate between $s_u$ and $i_u, i_e$, $s_e$ and $i_u, i_e$. & Persons$^{-1} \cdot $ Days$^{-1}$  \\ \hline
$\phi_{u,e}$ & Recovery rate for $i_u$, $i_e$ & Days$^{-1}$  \\ \hline
$\xi_{u,e}$ & Loss-of-immunity initiation rate for $r_u$, $r_e$ & Days$^{-1}$  \\ \hline
$\chi_u^{u,e}$, $\chi_e^{u,e}$ & Immunity boosting rates for $r_u$, $r_e$ & Persons$^{-1} \cdot $ Days$^{-1}$   \\ \hline
$\kappa_{u,e}$ & Loss-of-immunity rate for $w_u$, $w_e$ & Days$^{-1}$  \\ \hline
$\gamma_{u,e}$ & Mortality rate from disease, for $i_u$, $i_e$ & Days$^{-1}$  \\ \hline
$\alpha$ & Vaccination rate & Days$^{-1}$  \\ \hline
$\psi$ & Natality rate  & Days$^{-1}$   \\ \hline
$\rho$  & General mortality rate  & Days$^{-1}$   \\ \hline
\end{tabular}
\caption{Relevant variables, parameters, parameter names, and units for the system  \eqref{eqn:s_u}-\eqref{eqn:m}.}
  \label{tab:1DParametertTable} 
\end{center}
\end{table}






\section{Numerical results}

We show a brief demonstration of the model's capability by running it for a hypothetical scenario. We take parameter values as shown in Table \ref{tab:simulationValues}. We note that these values do not necessarily correspond to any particular outbreak values; they are designed to showcase the type of solution that the model \eqref{eqn:s_u}-\eqref{eqn:m}
can admit, in particular on a long term. We simulate the model over a two-year period. The model is implemented in MATLAB \cite{MATLAB:2019}. Since the the underlying equations are stiff, we utilize the ode23s solver.
\begin{table}
\begin{center}
\begin{tabular}{ |c|c|c| } 
\hline
Parameter/variable &  Units  &  Value   \\
\hline\hline

$s_{u_0, e_0}$ & Persons & 9999, 0  \\ \hline
$i_{u_0, e_0}$ &  Persons & 8, 0\\ \hline
$r_{u_0, e_0}$ & Persons & 0, 0\\ \hline
$w_{u_0, e_0}$  & Persons & 0, 0\\ \hline
$m$ & Persons & 0 \\ \hline
$\beta_u^{u,e}$ & Persons$^{-1} \cdot $ Days$^{-1}$ & 1.54e-5, 4e-5 \\ \hline
$\beta_e^{u,e}$ & Persons$^{-1} \cdot $ Days$^{-1}$ & 5.5e-5, 5.5e-5  \\ \hline
$\phi_{u,e}$ & Days$^{-1}$ & .13, .14  \\ \hline
$\xi_{u,e}$  & Days$^{-1}$ & .005, .005 \\ \hline
$\chi_u^{u,e}$ & Persons$^{-1} \cdot $ Days$^{-1}$  & 7.7.e-5, 2.e-4 \\ \hline
$\chi_e^{u,e}$ & Persons$^{-1} \cdot $ Days$^{-1}$  & 2.8e-4, 2.8e-4 \\ \hline
$\kappa_{u,e}$ & Days$^{-1}$ & .005, .005  \\ \hline
$\gamma_{u,e}$ & Days$^{-1}$ & .003, 4.4e-5  \\ \hline
$\alpha$ & Days$^{-1}$ & .002 \\ \hline
$\psi$ & Days$^{-1}$ & 5e-5   \\ \hline
$\rho$ &  Days$^{-1}$ & 5e-5  \\ \hline
\end{tabular}
\caption{Parameter values for the shown simulation.}
  \label{tab:simulationValues} 
\end{center}
\end{table}

\begin{figure}\centering
  \includegraphics[width=\textwidth]{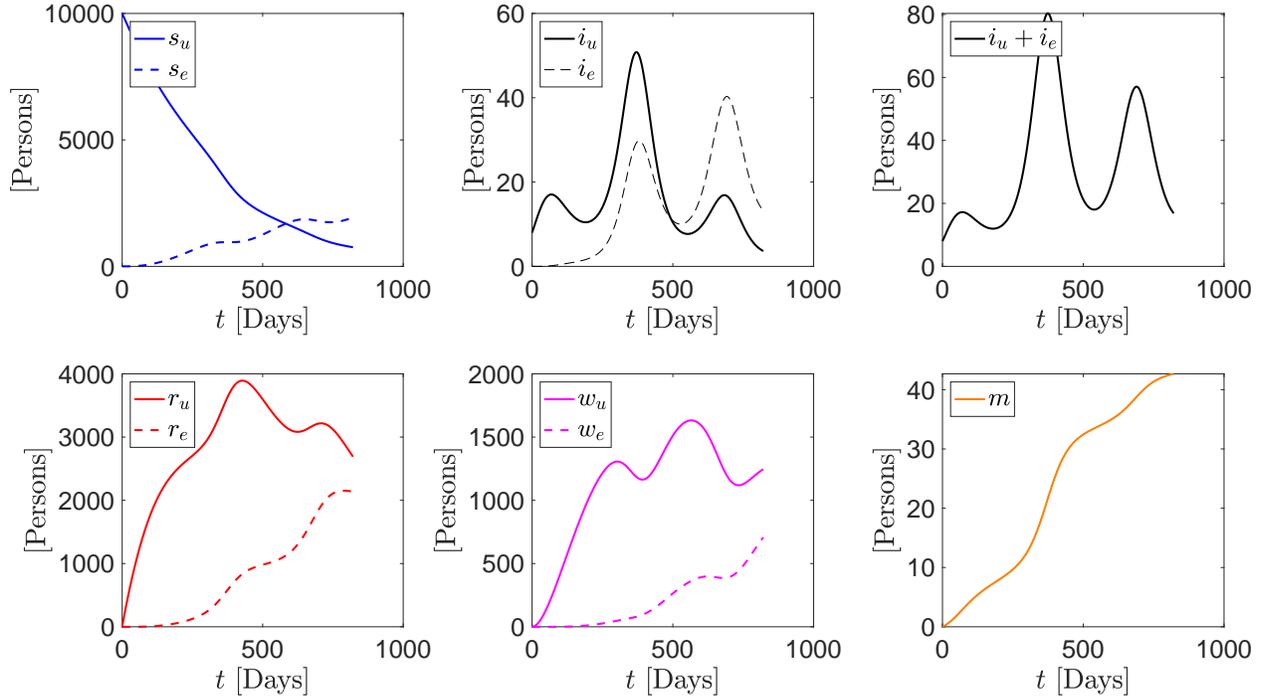}
 \caption{Simulation results with the model \eqref{eqn:s_u}-\eqref{eqn:m} with the parameters found in Table \ref{tab:1DParametertTable}. With constant parameter values, the model is able to reproduce the qualitative behavior of the COVID-19 pandemic.}\label{fig:simResults}
\end{figure}

The results of the simulation are shown in Fig. \ref{fig:simResults}. There are several important characteristics to note: the first is that, as shown in the second and third figures on the top row, the solution exhibits a periodic behavior, similar to the observed dynamic. The second is that, despite the total number of infections remaining high, in the third wave, we see a marked drop in mortality; this is due to larger percentages of the population having gained immunity, either through prior exposure to the disease, vaccination, or both.
\par
\begin{figure}\centering
  \includegraphics[width=\textwidth]{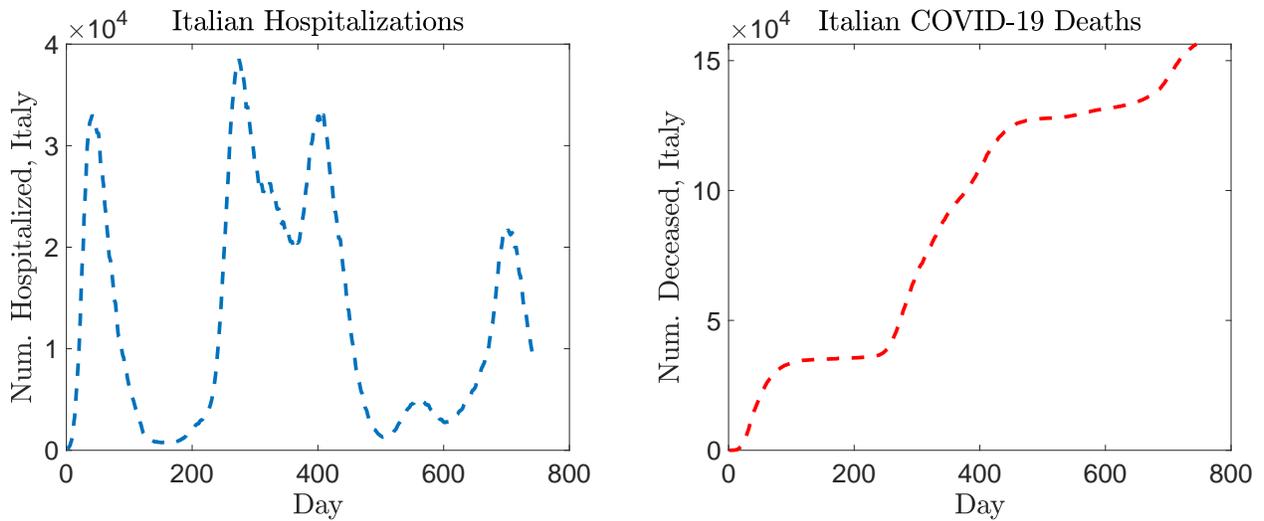}
 \caption{Number of hospitalized persons (left) and deaths (right) in Italy. We see that, when compared to Fig. \ref{fig:simResults}, many of the key features are preserved: the qualitative behavior of the infection and deceased curves show a good resemblance with the observed behavior of the COVID-19 pandemic. }\label{fig:italyComparison}
\end{figure}

 In Fig. \ref{fig:italyComparison}, we show the number of patients hospitalized with COVID-19 (left) and total deceased from COVID-19 (right) in Italy \cite{ItalianPC}. We note that large changes in testing protocols throughout the pandemic make a direct use of case-counts unreliable, and hence we consider hospitalizations, as this data is less subject to outside influence. We note that both the hospitalization and death data display the same general qualitative trends as in the simulation; the hospitalization data shows a clear periodic behavior, and the death curve shows intermittent periods of stalling and growth, with the overall trend flattening in time. We believe that, as in our model, this reduction in severe illness, despite continuing high levels of the disease, is due to the large amount of previously acquired immunity in the population. We may expect that, even in the presence of continued periodic surges, the mortality rate will remain low.
\par \textit{We stress that the agreement is, at this stage, only qualitative and very preliminary.} A proper parameter fitting is required in order for the model to produce meaningful predictions. However, the results shown indicate that, given such a parameter fitting, the proposed model is capable of naturally producing the complex periodic dynamics observed during COVID-19, even without reparameterization. As such dynamics are beyond the reach of most similar models employed for COVID-19, we believe that such a result, despite its preliminary nature, is significant.
A fine tuning of the parameters should include possible seasonality and lockdown dynamics.

\section{Conclusions and Perspectives}
In this brief note, we have introduced an SIRW2 model, related to the models shown in \cite{jardon2021geometric, van2000waning}. The proposed model is intended to describe a more sophisticated immune-system process than that found in most models: by the incorporation of immune boosting, and by stratification in terms of persons with and without previous exposure to the disease (either by vaccination or prior infection). We then demonstrated, through proof-of-concept simulations, that this modeling immunity may naturally recreate wave-like behaviors.
This is consistent with the long-term results of COVID-19 outbreak, 
without requiring time-dependent parameter adjustment. Qualitatively, we obtain good agreement with observed COVID-19 infection dynamics, providing strong evidence that, after a full parameter-fitting process, the proposed model may be well-suited for COVID-19, as well as other infectious diseases showing a similar behavior.
Much work has to be done for further developments. 
In particular, the theory of the model must be further developed, including the analysis of a suitable definition of $R_0$ and $R_t$, stability and equilibrium solutions, and bifurcation analyses. Rigorous derivations describing the model behavior (for instance, the oscillation period) in terms of the model parameters is also important. Most importantly, an application of the model on real-world data, incorporating a rigorous parameter fitting process, is necessary as a quantitative validation. 
Finally, this modeling framework can of course be developed even further via the incorporation of additional factors, such as time-lags, hospitalizations, differentiation between symptomatic and asymptomatic infections, and other important epidemiological considerations.

 \bibliographystyle{unsrt}  
\bibliography{references}

\begin{thebibliography}{10}

\bibitem{ItalianPC}
Protezione~Civile Italiana.
\newblock Covid-19.
\newblock \url{https://github.com/pcm-dpc/COVID-19}, 2020.

\bibitem{kermack1927contribution}
William~Ogilvy Kermack and Anderson~G McKendrick.
\newblock A contribution to the mathematical theory of epidemics.
\newblock {\em Proceedings of the royal society of london. Series A, Containing
  papers of a mathematical and physical character}, 115(772):700--721, 1927.

\bibitem{breda2012formulation}
Dimitri Breda, Odo Diekmann, WF~De~Graaf, A~Pugliese, and R~Vermiglio.
\newblock On the formulation of epidemic models (an appraisal of {Kermack and
  McKendrick}).
\newblock {\em Journal of biological dynamics}, 6(sup2):103--117, 2012.

\bibitem{hethcote2000mathematics}
Herbert~W Hethcote.
\newblock The mathematics of infectious diseases.
\newblock {\em SIAM review}, 42(4):599--653, 2000.

\bibitem{viguerie2021simulating}
Alex Viguerie, Guillermo Lorenzo, Ferdinando Auricchio, Davide Baroli,
  Thomas~JR Hughes, Alessia Patton, Alessandro Reali, Thomas~E Yankeelov, and
  Alessandro Veneziani.
\newblock Simulating the spread of {COVID-19} via a spatially-resolved
  susceptible--exposed--infected--recovered--deceased {(SEIRD)} model with
  heterogeneous diffusion.
\newblock {\em Applied Mathematics Letters}, 111:106617, 2021.

\bibitem{grave2021assessing}
Mal{\'u} Grave, Alex Viguerie, Gabriel~F Barros, Alessandro Reali, and
  Alvaro~LGA Coutinho.
\newblock Assessing the spatio-temporal spread of {COVID-19} via compartmental
  models with diffusion in {I}taly, {USA}, and {B}razil.
\newblock {\em Archives of Computational Methods in Engineering},
  28(6):4205--4223, 2021.

\bibitem{zhang2021integrated}
Sheng Zhang, Joan Ponce, Zhen Zhang, Guang Lin, and George Karniadakis.
\newblock An integrated framework for building trustworthy data-driven
  epidemiological models: {A}pplication to the {COVID-19} outbreak in {New York
  City}.
\newblock {\em PLoS computational biology}, 17(9):e1009334, 2021.

\bibitem{parolini2021suihter}
Nicola Parolini, Luca Dede’, Paola~F Antonietti, Giovanni Ardenghi, Andrea
  Manzoni, Edie Miglio, Andrea Pugliese, Marco Verani, and Alfio Quarteroni.
\newblock Suihter: A new mathematical model for {COVID-19}. application to the
  analysis of the second epidemic outbreak in {I}taly.
\newblock {\em Proceedings of the Royal Society A}, 477(2253):20210027, 2021.

\bibitem{bhouri2021covid}
Mohamed~Aziz Bhouri, Francisco~Sahli Costabal, Hanwen Wang, Kevin Linka,
  Mathias Peirlinck, Ellen Kuhl, and Paris Perdikaris.
\newblock {COVID-19} dynamics across the {US}: A deep learning study of human
  mobility and social behavior.
\newblock {\em Computer Methods in Applied Mechanics and Engineering},
  382:113891, 2021.

\bibitem{jardon2021geometric}
Hildeberto Jard{\'o}n-Kojakhmetov, Christian Kuehn, Andrea Pugliese, and Mattia
  Sensi.
\newblock A geometric analysis of the {SIS}, {SIRS} and {SIRWS} epidemiological
  models.
\newblock {\em Nonlinear Analysis: Real World Applications}, 58:103220, 2021.

\bibitem{anggriani2022mathematical}
Nursanti Anggriani, Meksianis~Z Ndii, Rika Amelia, Wahyu Suryaningrat, and
  Mochammad Andhika~Aji Pratama.
\newblock A mathematical {COVID-19} model considering asymptomatic and
  symptomatic classes with waning immunity.
\newblock {\em Alexandria Engineering Journal}, 61(1):113--124, 2022.

\bibitem{crellen2021dynamics}
Thomas Crellen, Li~Pi, Emma~L Davis, Timothy~M Pollington, Tim~CD Lucas,
  Diepreye Ayabina, Anna Borlase, Jaspreet Toor, Kiesha Prem, Graham~F Medley,
  et~al.
\newblock Dynamics of {SARS-CoV-2} with waning immunity in the {UK} population.
\newblock {\em Philosophical transactions of the royal society b},
  376(1829):20200274, 2021.

\bibitem{altmann2021waning}
Daniel~M Altmann and Rosemary~J Boyton.
\newblock Waning immunity to {SARS-CoV-2}: implications for vaccine booster
  strategies.
\newblock {\em The Lancet Respiratory Medicine}, 9(12):1356--1358, 2021.

\bibitem{dolgin2021covid}
Elie Dolgin et~al.
\newblock {COVID} vaccine immunity is waning-how much does that matter.
\newblock {\em Nature}, 597(7878):606--607, 2021.

\bibitem{van2000waning}
Michiel van Boven, Hester~E de~Melker, Joop~FP Schellekens, and Mirjam
  Kretzschmar.
\newblock Waning immunity and sub-clinical infection in an epidemic model:
  implications for pertussis in {The Netherlands}.
\newblock {\em Mathematical Biosciences}, 164(2):161--182, 2000.

\bibitem{laidlaw2022germinal}
Brian~J Laidlaw and Ali~H Ellebedy.
\newblock The germinal centre {B} cell response to {SARS-CoV}-2.
\newblock {\em Nature Reviews Immunology}, 22(1):7--18, 2022.

\bibitem{MATLAB:2019}
MATLAB.
\newblock {\em version R2019b}.
\newblock The MathWorks Inc., Natick, Massachusetts, 2019.

\end{thebibliography}

\end{document}